\documentclass[prb,showpacs,superscriptaddress,twocolumn]{revtex4-1}
\usepackage{amssymb,amsmath,color,graphicx,psfrag}

\begin{document}

\newcommand{\kv}[0]{\mathbf{k}}
\newcommand{\Rv}[0]{\mathbf{R}}
\newcommand{\Hv}[0]{\mathbf{H}}
\newcommand{\Mv}[0]{\mathbf{M}}
\newcommand{\Vv}[0]{\mathbf{V}}
\newcommand{\Uv}[0]{\mathbf{U}}
\newcommand{\rv}[0]{\mathbf{r}}
\newcommand{\gv}[0]{\mathbf{g}}
\newcommand{\al}[0]{\mathbf{a_{1}}}
\newcommand{\as}[0]{\mathbf{a_{2}}}
\newcommand{\K}[0]{\mathbf{K}}
\newcommand{\Kp}[0]{\mathbf{K'}}
\newcommand{\dkv}[0]{\delta\kv}
\newcommand{\dkx}[0]{\delta k_{x}}
\newcommand{\dky}[0]{\delta k_{y}}
\newcommand{\dk}[0]{\delta k}
\newcommand{\cv}[0]{\mathbf{c}}
\newcommand{\dv}[0]{\mathbf{d}}
\newcommand{\qv}[0]{\mathbf{q}}
\newcommand{\Rr}[0]{\Rv_{\rv}}
\newcommand{\Gv}[0]{\mathbf{G}}
\newcommand{\ev}[0]{\mathbf{e}}
\newcommand{\uu}[1]{\underline{\underline{#1}}}

\newcommand{\jav}[1]{#1}

\title{Dynamics of entanglement after exceptional quantum quench}

\author{\'Ad\'am B\'acsi}
\email{bacsi.adam@sze.hu}
\affiliation{MTA-BME Lend\"ulet Topology and Correlation Research Group,
Budapest University of Technology and Economics, 1521 Budapest, Hungary}
\affiliation{Department of Mathematics and Computational Sciences, Sz\'echenyi Istv\'an University, 9026 Gy\H or, Hungary}
\author{Bal\'azs D\'ora}
\affiliation{MTA-BME Lend\"ulet Topology and Correlation Research Group,
Budapest University of Technology and Economics, 1521 Budapest, Hungary}
\affiliation{Department of Theoretical Physics, Budapest University of Technology and Economics, Budapest, Hungary}
\date{\today}

\begin{abstract}
We investigate a quantum quench from a critical to an exceptional point. The initial state, prepared  in the ground state of a critical hermitian system,
is time evolved with a non-hermitian SSH model, tuned to its exceptional point.
The single particle density matrix exhibits supersonic modes and multiple light cones, characteristic to non-hermitian time evolution. These propagate with integer multiples
of the original Fermi velocity.
In the long time limit, the fermionic Green's function decays spatially as $1/x^2$,
in sharp contrast to the usual $1/x$ decay of non-interacting fermions.
The entanglement entropy is understood as if all these supersonic modes arise from independent quasiparticles (though they do not), traveling with the corresponding
supersonic light cone velocity. The entropy production rate decreases with time and develops plateaus during the time evolution, 
signaling the distinct velocities in the propagation of non-local
quantum correlations. At late times, the entanglement entropy saturates to a finite value, satisfying a volume law.
\end{abstract}

\maketitle

\section{Introduction}

Entanglement represents an important resource in quantum information science\cite{nielsen} and has become
 an indispensable tool for characterizing the ground states of quantum many-body systems\cite{eisert} and their time evolution.
In particular, entanglement plays a major role in investigating the thermodynamics of black holes \cite{srednicki} and (topological) order \cite{amico,pollmann2010}.
Its time evolution and dynamics can also be used to classify the properties of the underlying system\cite{aolita,znidaric2008,Calabrese_2005,calabrese_2018}.
While it has been thoroughly investigated in hermitian quantum systems, much less is known about its dynamics within the non-hermitian realm\cite{ashida18,alba,lei19,lee20}.

Non-hermitian quantum systems\cite{foatorres,ashidareview} have attracted a great deal attention recently, which are
natural extensions of conventional quantum mechanics to settings in which a
system interacts with the outside world, e.g. in the form of coupling
to a bath\cite{daley,carmichael,naghilo_exp_2019,zhou18}.
In this setting, the concept of an exceptional point\cite{heiss} (EP)
replaces the idea of a critical point\cite{sachdev}.
At an EP,  the complex spectrum becomes gapless when two (or more) complex eigenvalues and eigenstates coalesce and  no longer form a complete basis.
Due to this, the dynamics at EP is expected to be completely
distinct from what we are used to in hermitian quantum mechanics. 
For example, simple correlation function after quantum quench feature 
supersonic modes\cite{ashida18,dora2020}, propagating faster than the maximal velocity due to non-hermitian dynamics, thus violating the Lieb-Robinson\cite{liebrobinson} bound.
From this perspective, it looks natural to ask how the dynamics of entanglement and quantum information is affected in non-hermitian systems.

We focus on the fate of the ground state of critical one dimensional system, time evolved with a non-hermitian Hamiltonian containing an EP. 
For that, we use a non-hermitian variant\cite{lieu,longhi13} of the one-dimensional Su-Schrieffer-Heeger\cite{ssh} (SSH) model, supplemented with alternating imaginary on-site potential.
The equal time fermionic Green's function reveals two surprising features: it exhibits multiple light cones and supersonic modes, propagating with integer multiples of the maximal Fermi velocity
from the dispersion, and in the steady state, it decays spatially as $1/x^2$, in sharp contrast to the typical $1/x$ behaviour of non-interacting one dimensional critical systems.
The former is inherited also in the dynamics of subsystem entanglement entropy, which 
is expected to increase  with time\cite{Calabrese_2005,calabrese_2018}. On top of this, we find that 
the entropy production rate develops plateaus, separated by the timescales when a given supersonic mode
leaves the subsystem. The late time entanglement entropy obeys a volume law.
This can be qualitatively understood based on a simple picture, in which the supersonic modes are assigned to independent quasiparticles.

\section{Non-hermitian Quench protocol} 
\begin{figure}[h!]
\centering
\includegraphics[width=7cm]{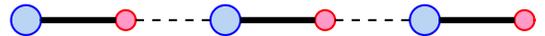}
\caption{Schematics of the non-hermitian SSH model after the quench. The small/big circles denote the $A$/$B$ sublattices, subject to balanced ($h$)
gain (red)/loss (blue), respectively. The thick solid/thin
dashed lines stand for the alternating hoppings, $J+\frac h2$ and $J-\frac h2$. The distance between two adjacent $A$s (or $B$s) is $a$.}
\label{fig:ssh}
\end{figure}
We consider the non-hermitian time evolution of the one-dimensional Su-Schrieffer-Heeger model, tuned to its EP\cite{heiss} following a 
sudden quench at $t=0$. Initially, the system is prepared in the ground state of the hermitian tight-binding Hamiltonian. 
\begin{gather}
H_0 = -J\sum_j \left( c_{Aj}^{+}c_{Bj-1}+c_{Bj-1}^+c_{Aj} + c_{Aj}^{+}c_{Bj} + c_{Bj}^{+}c_{Aj}\right),
\label{H0}
\end{gather}
where $J>0$ is the hopping amplitude. Although all atoms are equivalent in the initial model, we introduce $A$ and $B$ sublattices for later convenience.

By using Fourier transformation $c_{A/Bk}=\frac{1}{\sqrt{N}}\sum_j e^{ikx_j}c_{A/Bj}$ and introducing 
$d_{k+}=\frac{1}{\sqrt{2}}\left(c_{Ak}-e^{i\frac{ka}{2}}c_{Bk}\right)$ and $d_{k-}=\frac{1}{\sqrt{2}}\left(e^{-i\frac{ka}{2}}c_{Ak}+c_{Bk}\right)$, this
 Hamiltonian is diagonalized as
\begin{gather}
H_0=\sum_k \varepsilon(k)\left(d_{k+}^{+}d_{k+}  - d_{k-}^{+}d_{k-}\right),
\label{eq:H0}
\end{gather}
describing two bands $\varepsilon(k) = 2J\cos(ka/2)$ and $-\varepsilon(k)$. Here, $N$ denotes the number of unit cells and $a$ is the lattice constant, 
i.e., the distance between two adjacent $A$ atoms or adjacent $B$ atoms, Fig. \ref{fig:ssh}. 
The sum over $k$ runs over the Brillouin zone $(-\pi/a;\pi/a)$.
Note that the two bands touch each other at the edge of the Brillouin zone $k=\pm\pi/a$. Therefore, the system is critical
since there are low energy excitation\jav{s} around the touching point.
The initial state is half-filled, the lower band is completely filled while the upper band is empty.

At $t=0$, a sudden quench is performed in which an alternating hermitian hopping and alternating imaginary, non-hermitian site energies 
are switched on, representing a specific non-hermitian SSH model\cite{lieu,longhi13,qiu19} tuned to its EP, visualized in Fig. \ref{fig:ssh}. The Hamiltonian for $t>0$ reads
\begin{gather}
H=H_0 + \sum_{j}\left[ih \left(c_{Aj}^+ c_{Aj} - c_{Bj}^+c_{Bj}\right)+\right. \nonumber \\
\left. + \frac{h}{2}\left(c_{Aj}^+c_{Bj-1} + c_{Bj-1}^+ c_{Aj}\right) - \frac{h}{2}\left(c_{Bj}^+c_{Aj} + c_{Bj}^+c_{Aj}\right)\right].
\label{eq:fullham}
\end{gather}
The complex site energy stems from coupling to the environment\cite{longhi13,lieu} with alternating balanced particle gain and loss on sublattice $A$ and $B$.
Although the Hamiltonian becomes non-hermitian due to the complex site energy, the system 
fulfills $\mathcal{PT}$-symmetry~\footnote{The parity $\mathcal{P}$ is the spatial inversion to the midpoint of an $AB$ bond.} 
\cite{Bender_PRL_1998,mannheim_2013} and, hence, its spectrum remains real-valued for $2J>|h|$ as
\begin{gather}
E_{\pm}(k) = \pm E_0\cos\left(\frac{ka}{2}\right)
\label{spectrum}
\end{gather}
with $E_0 = \sqrt{4J^2-h^2}$. The spectrum hardly changes, only the original hopping amplitude $J$ gets renormalized.  
Consequently, the spectrum remains gapless  after the quench, and the $ka=\pm\pi$ points are exceptional points.
By linearizing the spectrum around the EPs, the Fermi velocity, which is the maximal velocity of propagation of excitations, becomes $v_F=E_0a/2$.
In spite of these mild looking changes, the resulting dynamics is completely different from that in  hermitian systems and the imaginary
alternating site energy in Eq. \eqref{eq:fullham} will play pivotal role, as we demonstrate later.

Our main goal is to determine how the entanglement entropy of a subsystem changes with time. The whole system is an infinitely long chain while
 the subsystem is a finite section of it. In order to evaluate the entanglement between the subsystem and the rest of the chain, we calculate 
the equal-time Green's function first. Based on Refs. \onlinecite{Peschel2009,herviou,chen20,chang20,maity}, the entanglement entropy is  derived directly from the Green's function.

\section{Equal-time Green's function after quantum quench}
Since the system consists of two sublattices, the Green's function has the structure of a $2\times 2$ matrix with the elements of
\begin{gather}
G_{XX'}(x_j,x_j';t)=\langle \Psi(t) | c_{Xj}c_{X'j'}^{+}|\Psi(t)\rangle,
\end{gather}
where $X$ and $X'$ stand for sublattice $A$ or $B$, $j$ and $j'$ index the unit cells.
 \jav{The time-dependent wavefunction $\Psi(t)$ is the solution of} the non-hermitian Schr\" odinger equation $i\partial_t|\Psi(t)\rangle=H|\Psi(t)\rangle$ with the ground state wavefunction of Eq. \eqref{H0} as the initial condition. 
Due to translational invariance, the Green's function depends only on $x_{j'}-x_{j}$  and factorizes to separate wavenumber channels as
\begin{gather}
\Gv(x_{j'}-x_j;t)=\frac{1}{N}\sum_k e^{ik(x_{j'}-x_{j})}\Gv_k(t),
\label{eq:greenFT}
\end{gather}
where $\Gv_k(t)$ is the time-dependent Green's matrix in a specific momentum channel.
Using the time-dependent Schr\" odinger equation, the elements of the $\Gv_k(t)$ matrix can be obtained analytically using the equation of motion method (see Appendix).
We note that although the Hamiltonian itself is quadratic in terms of fermionic creation and annihilation operators, the resulting hierarchy of equations of motion becomes
more involved due to the non-hermitian Hamiltonian as for  the hermitian case.

An interesting feature of non-hermitian dynamics is that
 the norm of the wavefunction does not necessarily remain unity for the whole time evolution. In our case, the norm square depends
 on time as $\mathcal{N}_k(t)=1 + (1+C_k)A_k(t)$
in a given $k$ momentum channel, which allows us to write norm square of the many-body wavefunction 
as $\langle \Psi(t)|\Psi(t)\rangle = \prod_{k}\mathcal{N}_k(t)$. To retain the norm, the wavefunction must be normalized\cite{ashidareview,graefe2008,carmichael,ashida18}
 as $|\Psi(t)\rangle/\sqrt{\langle\Psi(t)|\Psi(t)\rangle}$ which leads to the normalized Green's function
\begin{gather}
\tilde{\Gv}_k(t) = \frac{\Gv_k(t)}{\mathcal{N}_k(t)} = 
\Gv_{k}(0) 
+\frac{\sqrt{C_k}}{\mathcal{N}_k(t)}\times\nonumber\\
\times\left[\begin{array}{cc} 
-B_k(t) & e^{i\frac{ka}{2}}A_k(t)\left(\sqrt{C_k} + i\frac{2J}{h}\right) \\ 
e^{-i\frac{ka}{2}}A_k(t)\left(\sqrt{C_k} - i\frac{2J}{h}\right) & B_k(t) \end{array}\right],
\label{eq:Gofk}
\end{gather}
where $A_k(t)=\frac{h^2}{2E_0^2}\left(1-\cos(2E_k t)\right)$, $B_k(t) = \frac{h}{2E_0}\sin(2E_kt)$, $C_k = \frac{1+\sin(ka/2)}{1 - \sin(ka/2)}$
and
\begin{gather}
\Gv_{k}(0) =\frac{1}{2}\left[\begin{array}{cc} 1 & -e^{\frac{ika}{2}} \\ -e^{-\frac{ika}{2}} & 1 \end{array}\right]
\label{eq:g0}
\end{gather}
is the initial Green's matrix, corresponding to the simple, hermitian tight-binding initial Hamiltonian.

The Fourier transform of the initial Green's function is calculated as
\begin{gather}
\Gv(x; t=0) =\left[ \begin{array}{cc} \frac{1}{2} \delta_{x,0} & 
\frac{1}{2\pi}\frac{(-1)^{x/a + 1}}{x/a + \frac{1}{2}} \\
\frac{1}{2\pi}\frac{(-1)^{x/a + 1}}{-x/a + \frac{1}{2}} & \frac{1}{2} \delta_{x,0} \end{array}\right],
\end{gather}
obeying the well known $1/x$ decay of free fermions in one dimension\cite{giamarchi}. Note that $x$ can only be an integer multiple of $a$.
Here, only the off-diagonal component contributes for finite $x$, describing correlations between distinct sublattices, or on the original one atom per unit cell basis, between
lattice sites with odd distance between them.

\begin{figure}[h!]
\centering
\psfrag{x}[t][][1][0]{$x/2v_F t$}
\psfrag{y}[b][t][1][0]{$|\tilde\Gv_{AB}(x=1000a,t)|$}
\psfrag{xx}[t][][1][0]{$x/a$}
\psfrag{yy}[b][t][1][0]{$|\tilde{\Gv}_{AB}^\infty(x)|$}
\includegraphics[width=7cm]{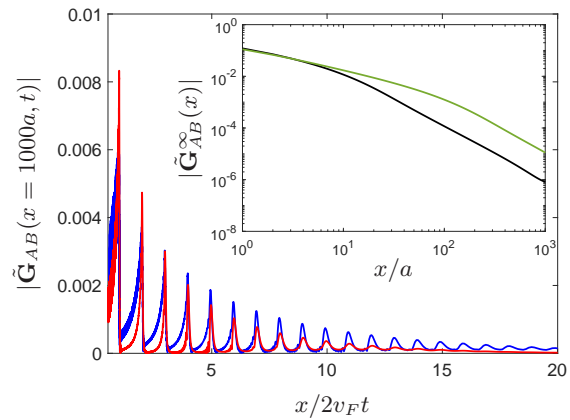}
\caption{Time dependence of the absolute value of the off-diagonal Green's function is plotted for $x/a=1000$ with $h/J=1$ (blue) and 0.5 (red). The other, $AA$, $BA$ and $BB$ components 
are indistinguishable from $AB$ on this scale.
Each peak at integer $x/2v_Ft$ is identified as a supersonic mode.
The inset depicts the time averaged off diagonal Green's function in the steady state for $h=0.1$ (black) and 0.01 (green), the diagonal components vanish for $x\neq 0$. 
The spatial decay changes from $1/x$ to $1/x^2$
with increasing $x$, the crossover occurs at $x\sim J/h$.}
\label{fig:greenfull}
\end{figure}

The time-dependent part of Eq. \eqref{eq:Gofk} is evaluated numerically and is shown in Fig. \ref{fig:greenfull} for some representative values of $h$.
The Green's function exhibits several peaks and multiple light cones\cite{groha}. By studying the time evolution, we find that 
the peaks travel with even multiples of the Fermi velocity $v_F$ around the EP. The different peaks, therefore, correspond to different supersonic modes and different light cones,
propagating with distinct velocities. This phenomenon is similar to that found in Refs. \onlinecite{ashida18,dora2020}. 
Emergence of the supersonic modes is the consequence of the non-hermitian dynamics and renormalization of the wavefunction. Since $[H,H^+]\neq 0$ is the non-hermitian case,
an effective long-range Hamiltonian is generated during the time evolution, for which the conventional Lieb-Robinson bound\cite{liebrobinson} 
does not apply, and excitations can propagate faster
than the effective Fermi velocity, $v_F$.

The non-hermitian time evolution, we consider, can originate from an open quantum system Lindbladian dynamics by restricting our attention to the subspace with no quantum jumps\cite{daley,ashida18}.
By allowing for quantum jumps and the full Lindblad time evolution, no signatures of supersonic modes have been found (see Appendix).
 
To describe the steady state, we study the time average of the Green's matrix as
$\tilde{\Gv}_k^\infty=\lim_{t\rightarrow \infty}\frac{1}{t}\int_0^t \tilde{\Gv}_k(t')\,\mathrm{d}t'$
which is obtained analytically as
\begin{gather}
\tilde{\Gv}_k^\infty = \Gv_k(0) + \frac{\sqrt{C_k}}{1+C_k}\left(1-\frac{E_0}{\sqrt{4J^2+h^2 C_k}}\right) \times\nonumber \\
\times\left[\begin{array}{cc} 0 & e^{i\frac{ka}{2}}\left(\sqrt{C_k} + i\frac{2J}{h}\right) \\
e^{-i\frac{ka}{2}}\left(\sqrt{C_k} - i\frac{2J}{h}\right) & 0 \end{array}\right].
\end{gather}
Similarly to the initial state, only off-diagonal correlations are present after Fourier transformation. Surprisingly, we find that the long distance asymptotics of the steady
state Green's function crosses over from  $1/x$ to $E_0/x^2$ with increasing $x$ at around $x\sim J/h$. Even though the elementary excitations are fermionic, the long distance
decay of the single particle density matrix changes significantly compared to those expected from free fermions\cite{giamarchi}. \jav{This 
 follows from the quench, which heats up the system from its zero temperature initial state. We have checked numerically that the $1/x^2$ decay in the steady state also appears
in 
quenches to gapped hermitian and non-hermitian SSH models, i.e. Eq. \eqref{eq:fullham} with unequal alternating hopping and imaginary site energies (see Appendix).}

\section{Time evolution of entanglement}

The entanglement entropy is obtained following Refs. \onlinecite{Peschel2009,herviou,ashida18}. 
To calculate the entanglement of a subsystem with the rest of the whole system, we obtain the correlation matrix 
$C_{jX,j'X'}(t)=\langle \Psi(t)|c_{Xj}^+ c_{X'j'}|\Psi(t)\rangle / \langle \Psi(t)|\Psi(t)\rangle$ in the subsystem 
which is a finite section of the whole chain. Here, $j,j'$ are the unit cell indices of the subsystem and $X,X'$ are $A$ or $B$. 
Note that $\mathbf{C}$ is closely related to the Green's function by $C_{jX,j'X'}(t) = \delta_{jj'}\delta_{XX'} - \tilde{G}_{X'X}(x_j-x_{j'};t)$. 
Given the matrix $\mathbf{C}$, we obtain its eigenvalues $\zeta_l(t)$ numerically, and based on Ref. \onlinecite{Peschel2009}, we calculate the entanglement entropy as
\begin{gather}
S(t) = -\sum_{l} \left[\left(1-\zeta_l\right)\ln\left(1-\zeta_l\right) + \zeta_l \ln\zeta_l\right].
\end{gather}
Our goal is determine the entanglement entropy for a subsystem of length $L$, consisting of $n$ unit cells with $L=na$.
Note that the subsystem contains $2n$ atomic sites and, hence, $\mathbf{C}$ is a $2n\times 2n$ matrix with $2n$ eigenvalues.

\begin{figure}[h!]
\centering
\psfrag{x}[t][][1][0]{$tJ$}
\psfrag{y}[b][t][1][0]{$S(t)$}
\includegraphics[width=6cm]{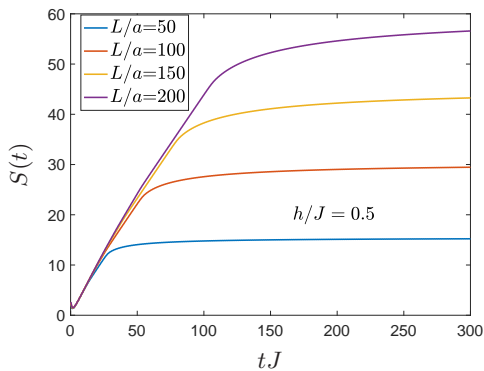}
\caption{The time evolution of the entanglement entropy is shown for several subsystem sizes. The initial linear growth in time, the gradual slope changes and the saturation for late time, 
obeying a volume law, is apparent.}
\label{fig:entropy0}
\end{figure}

\begin{figure}[h!]
\centering
\psfrag{x}[t][][1][0]{$2v_F t/L$}
\psfrag{y}[b][t][1][0]{$S'(t)/J$}
\psfrag{text}[][][0.9][0]{$L/a=1000$, $h/J=0.4$}
\includegraphics[width=7cm]{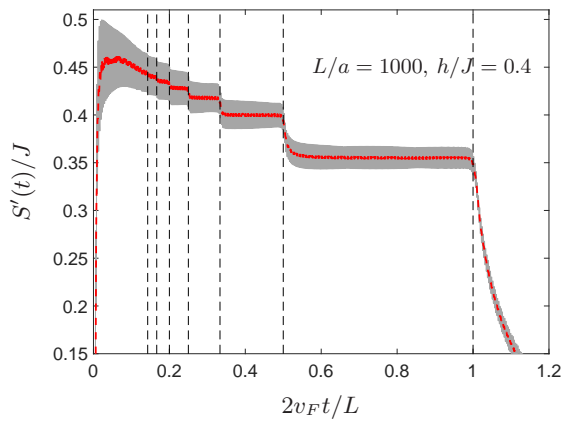}
\caption{The time evolution of the entropy production rate (grey line) is shown for a typical quench, the moving average (red dashed line) is also depicted to clearly visualize the plateaus.
The vertical black dashed lines indicate the expected plateau positions from the velocity of the supersonic modes for several $2v_Ft/L=1/m$ with $m=1$\dots 7.}
\label{fig:entropy}
\end{figure}


From the behaviour of the Green's function, the supersonic modes
are expected to show up also in the entanglement entropy. In order to reveal their presence, we focus on the entanglement entropy production rate, i.e. 
$S'(t)\equiv\partial_t S(t)$.
As a rough approximation, we can think of the supersonic modes as arising from distinct, independent quasiparticles (though there are only one type of fermionic excitations
in our system as exemplified by Eq. \eqref{spectrum}). Then, as a simple extension of Ref. \onlinecite{Calabrese_2005}, the ensuing entanglement entropy is the sum of independent contributions from the various quasiparticles, each propagating with its own sound velocity $v_m$,
which is
integer multiple of $v_F$, i.e. $v_m=mv_F$ with $m>0$ integer. 
The respective entanglement entropies are expected to rise linearly with time\cite{fagotti} with the corresponding entropy production rate, $\gamma_m$
and saturate to a steady state, volume law value when the light cone reaches the subsystem length, $L=2 v_m t$. Therefore, to a good approximation, we write
\begin{gather}
S(t)\approx \sum\limits_{m=1}^\infty\gamma_m t\Theta(L-2v_mt)+\frac{\gamma_m L}{2 v_m}\Theta(2v_mt-L),
\end{gather}
and the entropy production rate is
\begin{gather}
S'(t)\approx  \sum_{m=1}^{\infty} \gamma_m \cdot \Theta(L-2v_mt),
\label{eq:sfit}
\end{gather}
where $\Theta(x)$ is the Heaviside function. A direct evaluation of $\gamma_m$, similarly to Ref. \onlinecite{Alba2017} does not seem applicable since
for each $k$ mode, there is only a single fermionic branch in Eq. \eqref{spectrum}, the supersonic modes follow from the non-hermitian and non-unitary time evolution and \emph{not}
from different species of stable quasiparticles.

In Figs. \ref{fig:entropy0} and \ref{fig:entropy}, we show the numerical results for the entanglement entropy and its production rate, in perfect agreement with the above expectations.
We find that for a  subsystem of length $L$, distinct plateaus appear in the intervals $L/m<2 v_F t<L/(m-1)$ where $m$ is a positive integer. 
The heights of the individual plateaus  increase with $m$ but the jump from between adjacent  plateaus from $m-1$ to $m$  decreases with $m$. 
The times when the jumps occur are the necessary period to a supersonic mode to travel through the subsystem. 

\section{Conclusion}
We investigated the fate of a quantum critical wavefunction, upon quantum quenching with a non-hermitian Hamiltonian, containing an exceptional point.
The fermionic Green's function exhibits  multiple light cones and supersonic modes, propagating with integer multiples of the Fermi velocity of the elementary excitations.
In the steady state, it decays in a power law fashion as $1/x^2$ asymptotically, in spite of the non-interacting fermionic nature of the quasiparticles.
The entanglement between a finite subsystem of length $L$ with the rest of the infinitely long system is inspected.
The ensuing entanglement entropy is evaluated numerically, and reveals the presence of supersonic modes.
The entropy production rate decreases with time and develops plateaus during the time evolution,
signaling the distinct velocities in the propagation of non-local
quantum correlations. In the steady state, the entanglement entropy saturates to a finite value, satisfying a volume law.
All these features are qualitatively understood using a simple model, in which the supersonic modes are assigned to independent quasiparticles.

\begin{acknowledgments}
Illuminating discussions with Pasquale Calabrese, Ching Hua Lee and Frank Pollmann are gratefully acknowledged.
This research is supported by the National Research, Development and Innovation Office - NKFIH   within the Quantum Technology National Excellence Program (Project No.
      2017-1.2.1-NKP-2017-00001), K119442, K134437 and by the BME-Nanotechnology FIKP grant (BME FIKP-NAT).
\end{acknowledgments}

\bibliographystyle{apsrev}
\bibliography{sshquench,wboson1}

\begin{thebibliography}{10}
\expandafter\ifx\csname bibnamefont\endcsname\relax
  \def\bibnamefont#1{#1}\fi
\expandafter\ifx\csname bibfnamefont\endcsname\relax
  \def\bibfnamefont#1{#1}\fi
\expandafter\ifx\csname url\endcsname\relax
  \def\url#1{\texttt{#1}}\fi
\expandafter\ifx\csname urlprefix\endcsname\relax\def\urlprefix{URL }\fi
\providecommand{\bibinfo}[2]{#2}
\providecommand{\eprint}[2][]{\url{#2}}

\bibitem{nielsen}
\bibinfo{author}{\bibfnamefont{M.}~\bibnamefont{Nielsen}} \bibnamefont{and}
  \bibinfo{author}{\bibfnamefont{I.}~\bibnamefont{Chuang}},
  \emph{\bibinfo{title}{Quantum Computation and Quantum Information}}
  (\bibinfo{publisher}{Cambridge University Press},
  \bibinfo{address}{Cambridge}, \bibinfo{year}{2000}).

\bibitem{eisert}
\bibinfo{author}{\bibfnamefont{J.}~\bibnamefont{Eisert}},
  \bibinfo{author}{\bibfnamefont{M.}~\bibnamefont{Cramer}}, \bibnamefont{and}
  \bibinfo{author}{\bibfnamefont{M.~B.} \bibnamefont{Plenio}},
  \emph{\bibinfo{title}{\textit{Colloquium} : Area laws for the entanglement
  entropy}}, \bibinfo{journal}{Rev. Mod. Phys.} \textbf{\bibinfo{volume}{82}},
  \bibinfo{pages}{277} (\bibinfo{year}{2010}).

\bibitem{srednicki}
\bibinfo{author}{\bibfnamefont{M.}~\bibnamefont{Srednicki}},
  \emph{\bibinfo{title}{Entropy and area}}, \bibinfo{journal}{Phys. Rev. Lett.}
  \textbf{\bibinfo{volume}{71}}, \bibinfo{pages}{666} (\bibinfo{year}{1993}).

\bibitem{amico}
\bibinfo{author}{\bibfnamefont{L.}~\bibnamefont{Amico}},
  \bibinfo{author}{\bibfnamefont{R.}~\bibnamefont{Fazio}},
  \bibinfo{author}{\bibfnamefont{A.}~\bibnamefont{Osterloh}}, \bibnamefont{and}
  \bibinfo{author}{\bibfnamefont{V.}~\bibnamefont{Vedral}},
  \emph{\bibinfo{title}{Entanglement in many-body systems}},
  \bibinfo{journal}{Rev. Mod. Phys.} \textbf{\bibinfo{volume}{80}},
  \bibinfo{pages}{517} (\bibinfo{year}{2008}).

\bibitem{pollmann2010}
\bibinfo{author}{\bibfnamefont{F.}~\bibnamefont{Pollmann}},
  \bibinfo{author}{\bibfnamefont{A.~M.} \bibnamefont{Turner}},
  \bibinfo{author}{\bibfnamefont{E.}~\bibnamefont{Berg}}, \bibnamefont{and}
  \bibinfo{author}{\bibfnamefont{M.}~\bibnamefont{Oshikawa}},
  \emph{\bibinfo{title}{Entanglement spectrum of a topological phase in one
  dimension}}, \bibinfo{journal}{Phys. Rev. B} \textbf{\bibinfo{volume}{81}},
  \bibinfo{pages}{064439} (\bibinfo{year}{2010}).

\bibitem{aolita}
\bibinfo{author}{\bibfnamefont{L.}~\bibnamefont{Aolita}},
  \bibinfo{author}{\bibfnamefont{F.}~\bibnamefont{de~Melo}}, \bibnamefont{and}
  \bibinfo{author}{\bibfnamefont{L.}~\bibnamefont{Davidovich}},
  \emph{\bibinfo{title}{Open-system dynamics of entanglement:a key issues
  review}}, \bibinfo{journal}{Reports on Progress in Physics}
  \textbf{\bibinfo{volume}{78}}(\bibinfo{number}{4}), \bibinfo{pages}{042001}
  (\bibinfo{year}{2015}).

\bibitem{znidaric2008}
\bibinfo{author}{\bibfnamefont{M.}~\bibnamefont{Znidaric}},
  \bibinfo{author}{\bibfnamefont{T.}~\bibnamefont{Prosen}}, \bibnamefont{and}
  \bibinfo{author}{\bibfnamefont{P.}~\bibnamefont{Prelovsek}},
  \emph{\bibinfo{title}{Many-body localization in the heisenberg $xxz$ magnet
  in a random field}}, \bibinfo{journal}{Phys. Rev. B}
  \textbf{\bibinfo{volume}{77}}, \bibinfo{pages}{064426}
  (\bibinfo{year}{2008}).

\bibitem{Calabrese_2005}
\bibinfo{author}{\bibfnamefont{P.}~\bibnamefont{Calabrese}} \bibnamefont{and}
  \bibinfo{author}{\bibfnamefont{J.}~\bibnamefont{Cardy}},
  \emph{\bibinfo{title}{Evolution of entanglement entropy in one-dimensional
  systems}}, \bibinfo{journal}{Journal of Statistical Mechanics: Theory and
  Experiment} \textbf{\bibinfo{volume}{2005}}(\bibinfo{number}{04}),
  \bibinfo{pages}{P04010} (\bibinfo{year}{2005}).

\bibitem{calabrese_2018}
\bibinfo{author}{\bibfnamefont{V.}~\bibnamefont{Alba}} \bibnamefont{and}
  \bibinfo{author}{\bibfnamefont{P.}~\bibnamefont{Calabrese}},
  \emph{\bibinfo{title}{{Entanglement dynamics after quantum quenches in
  generic integrable systems}}}, \bibinfo{journal}{SciPost Phys.}
  \textbf{\bibinfo{volume}{4}}, \bibinfo{pages}{17} (\bibinfo{year}{2018}).

\bibitem{ashida18}
\bibinfo{author}{\bibfnamefont{Y.}~\bibnamefont{Ashida}} \bibnamefont{and}
  \bibinfo{author}{\bibfnamefont{M.}~\bibnamefont{Ueda}},
  \emph{\bibinfo{title}{Full-counting many-particle dynamics: Nonlocal and
  chiral propagation of correlations}}, \bibinfo{journal}{Phys. Rev. Lett.}
  \textbf{\bibinfo{volume}{120}}, \bibinfo{pages}{185301}
  (\bibinfo{year}{2018}).

\bibitem{alba}
\bibinfo{author}{\bibfnamefont{V.}~\bibnamefont{Alba}} \bibnamefont{and}
  \bibinfo{author}{\bibfnamefont{F.}~\bibnamefont{Carollo}},
  \emph{\bibinfo{title}{Spreading of correlations in markovian open quantum
  systems}}, \bibinfo{note}{{a}rXiv:2002.09527}.

\bibitem{lei19}
\bibinfo{author}{\bibfnamefont{L.}~\bibnamefont{Xiao}},
  \bibinfo{author}{\bibfnamefont{K.}~\bibnamefont{Wang}},
  \bibinfo{author}{\bibfnamefont{X.}~\bibnamefont{Zhan}},
  \bibinfo{author}{\bibfnamefont{Z.}~\bibnamefont{Bian}},
  \bibinfo{author}{\bibfnamefont{K.}~\bibnamefont{Kawabata}},
  \bibinfo{author}{\bibfnamefont{M.}~\bibnamefont{Ueda}},
  \bibinfo{author}{\bibfnamefont{W.}~\bibnamefont{Yi}}, \bibnamefont{and}
  \bibinfo{author}{\bibfnamefont{P.}~\bibnamefont{Xue}},
  \emph{\bibinfo{title}{Observation of critical phenomena in
  parity-time-symmetric quantum dynamics}}, \bibinfo{journal}{Phys. Rev. Lett.}
  \textbf{\bibinfo{volume}{123}}, \bibinfo{pages}{230401}
  (\bibinfo{year}{2019}).

\bibitem{lee20}
\bibinfo{author}{\bibfnamefont{C.~H.} \bibnamefont{Lee}},
  \emph{\bibinfo{title}{Exceptional boundary states and negative entanglement
  entropy}}, \bibinfo{note}{{a}rXiv:2011.09505}.

\bibitem{foatorres}
\bibinfo{author}{\bibfnamefont{L.~E. F.~F.} \bibnamefont{Torres}},
  \emph{\bibinfo{title}{Perspective on topological states of non-hermitian
  lattices}}, \bibinfo{journal}{Journal of Physics: Materials}
  \textbf{\bibinfo{volume}{3}}(\bibinfo{number}{1}), \bibinfo{pages}{014002}
  (\bibinfo{year}{2019}).

\bibitem{ashidareview}
\bibinfo{author}{\bibfnamefont{Y.}~\bibnamefont{Ashida}},
  \bibinfo{author}{\bibfnamefont{Z.}~\bibnamefont{Gong}}, \bibnamefont{and}
  \bibinfo{author}{\bibfnamefont{M.}~\bibnamefont{Ueda}},
  \emph{\bibinfo{title}{Non-hermitian physics}},
  \bibinfo{note}{{a}rXiv:2006.01837}.

\bibitem{daley}
\bibinfo{author}{\bibfnamefont{A.~J.} \bibnamefont{Daley}},
  \emph{\bibinfo{title}{Quantum trajectories and open many-body quantum
  systems}}, \bibinfo{journal}{Advances in Physics}
  \textbf{\bibinfo{volume}{63}}, \bibinfo{pages}{77} (\bibinfo{year}{2014}).

\bibitem{carmichael}
\bibinfo{author}{\bibfnamefont{H.}~\bibnamefont{Carmichael}},
  \emph{\bibinfo{title}{An Open Systems Approach to Quantum Optics}}
  (\bibinfo{publisher}{Springer-Verlag}, \bibinfo{address}{Berlin},
  \bibinfo{year}{1993}).

\bibitem{naghilo_exp_2019}
\bibinfo{author}{\bibfnamefont{M.}~\bibnamefont{Naghiloo}},
  \bibinfo{author}{\bibfnamefont{M.}~\bibnamefont{Abbasi}},
  \bibinfo{author}{\bibfnamefont{Y.~N.} \bibnamefont{Joglekar}},
  \bibnamefont{and} \bibinfo{author}{\bibfnamefont{K.~W.} \bibnamefont{Murch}},
  \emph{\bibinfo{title}{Quantum state tomography across the exceptional point
  in a single dissipative qubit}}, \bibinfo{journal}{Nature Physics}
  \textbf{\bibinfo{volume}{15}}, \bibinfo{pages}{1232} (\bibinfo{year}{2019}).

\bibitem{zhou18}
\bibinfo{author}{\bibfnamefont{L.}~\bibnamefont{Zhou}},
  \bibinfo{author}{\bibfnamefont{Q.-h.} \bibnamefont{Wang}},
  \bibinfo{author}{\bibfnamefont{H.}~\bibnamefont{Wang}}, \bibnamefont{and}
  \bibinfo{author}{\bibfnamefont{J.}~\bibnamefont{Gong}},
  \emph{\bibinfo{title}{Dynamical quantum phase transitions in non-hermitian
  lattices}}, \bibinfo{journal}{Phys. Rev. A} \textbf{\bibinfo{volume}{98}},
  \bibinfo{pages}{022129} (\bibinfo{year}{2018}).

\bibitem{heiss}
\bibinfo{author}{\bibfnamefont{W.~D.} \bibnamefont{Heiss}},
  \emph{\bibinfo{title}{The physics of exceptional points}},
  \bibinfo{journal}{J. Phys. A: Math. Theor.} \textbf{\bibinfo{volume}{45}},
  \bibinfo{pages}{444016} (\bibinfo{year}{2012}).

\bibitem{sachdev}
\bibinfo{author}{\bibfnamefont{S.}~\bibnamefont{Sachdev}},
  \emph{\bibinfo{title}{Quantum Phase Transitions}}
  (\bibinfo{publisher}{Cambridge Univ. Press}, \bibinfo{address}{Cambridge},
  \bibinfo{year}{1999}).

\bibitem{dora2020}
\bibinfo{author}{\bibfnamefont{B.}~\bibnamefont{D\'ora}} \bibnamefont{and}
  \bibinfo{author}{\bibfnamefont{C.~P.} \bibnamefont{Moca}},
  \emph{\bibinfo{title}{Quantum quench in $\mathcal{P}\mathcal{T}$-symmetric
  luttinger liquid}}, \bibinfo{journal}{Phys. Rev. Lett.}
  \textbf{\bibinfo{volume}{124}}, \bibinfo{pages}{136802}
  (\bibinfo{year}{2020}).

\bibitem{liebrobinson}
\bibinfo{author}{\bibfnamefont{E.}~\bibnamefont{Lieb}} \bibnamefont{and}
  \bibinfo{author}{\bibfnamefont{D.}~\bibnamefont{Robinson}},
  \emph{\bibinfo{title}{The finite group velocity of quantum spin systems}},
  \bibinfo{journal}{Commun. Math. Phys.} \textbf{\bibinfo{volume}{28}},
  \bibinfo{pages}{251} (\bibinfo{year}{1972}).

\bibitem{lieu}
\bibinfo{author}{\bibfnamefont{S.}~\bibnamefont{Lieu}},
  \emph{\bibinfo{title}{Topological phases in the non-hermitian
  su-schrieffer-heeger model}}, \bibinfo{journal}{Phys. Rev. B}
  \textbf{\bibinfo{volume}{97}}, \bibinfo{pages}{045106}
  (\bibinfo{year}{2018}).

\bibitem{longhi13}
\bibinfo{author}{\bibfnamefont{S.}~\bibnamefont{Longhi}},
  \emph{\bibinfo{title}{Convective and absolute $\mathcal{PT}$-symmetry
  breaking in tight-binding lattices}}, \bibinfo{journal}{Phys. Rev. A}
  \textbf{\bibinfo{volume}{88}}, \bibinfo{pages}{052102}
  (\bibinfo{year}{2013}).

\bibitem{ssh}
\bibinfo{author}{\bibfnamefont{W.~P.} \bibnamefont{Su}},
  \bibinfo{author}{\bibfnamefont{J.~R.} \bibnamefont{Schrieffer}},
  \bibnamefont{and} \bibinfo{author}{\bibfnamefont{A.~J.}
  \bibnamefont{Heeger}}, \emph{\bibinfo{title}{Solitons in polyacetylene}},
  \bibinfo{journal}{Phys. Rev. Lett.} \textbf{\bibinfo{volume}{42}},
  \bibinfo{pages}{1698} (\bibinfo{year}{1979}).

\bibitem{qiu19}
\bibinfo{author}{\bibfnamefont{X.}~\bibnamefont{Qiu}},
  \bibinfo{author}{\bibfnamefont{T.-S.} \bibnamefont{Deng}},
  \bibinfo{author}{\bibfnamefont{Y.}~\bibnamefont{Hu}},
  \bibinfo{author}{\bibfnamefont{P.}~\bibnamefont{Xue}}, \bibnamefont{and}
  \bibinfo{author}{\bibfnamefont{W.}~\bibnamefont{Yi}},
  \emph{\bibinfo{title}{Fixed points and dynamic topological phenomena in a
  parity-time-symmetric quantum quench}}, \bibinfo{journal}{iScience}
  \textbf{\bibinfo{volume}{20}}, \bibinfo{pages}{392 } (\bibinfo{year}{2019}).

\bibitem{Note1}
\bibinfo{note}{The parity $\protect \mathcal {P}$ is the spatial inversion to
  the midpoint of an $AB$ bond.}

\bibitem{Bender_PRL_1998}
\bibinfo{author}{\bibfnamefont{C.~M.} \bibnamefont{Bender}} \bibnamefont{and}
  \bibinfo{author}{\bibfnamefont{S.}~\bibnamefont{Boettcher}},
  \emph{\bibinfo{title}{Real spectra in non-hermitian hamiltonians having
  $\mathcal{PT}$ symmetry}}, \bibinfo{journal}{Phys. Rev. Lett.}
  \textbf{\bibinfo{volume}{80}}, \bibinfo{pages}{5243} (\bibinfo{year}{1998}).

\bibitem{mannheim_2013}
\bibinfo{author}{\bibfnamefont{P.~D.} \bibnamefont{Mannheim}},
  \emph{\bibinfo{title}{$\mathcal{PT}$ symmetry as a necessary and sufficient
  condition for unitary time evolution}}, \bibinfo{journal}{Philosophical
  Transactions of the Royal Society A: Mathematical, Physical and Engineering
  Sciences} \textbf{\bibinfo{volume}{371}}(\bibinfo{number}{1989}),
  \bibinfo{pages}{20120060} (\bibinfo{year}{2013}).

\bibitem{Peschel2009}
\bibinfo{author}{\bibfnamefont{I.}~\bibnamefont{Peschel}} \bibnamefont{and}
  \bibinfo{author}{\bibfnamefont{V.}~\bibnamefont{Eisler}},
  \emph{\bibinfo{title}{Reduced density matrices and entanglement entropy in
  free lattice models}}, \bibinfo{journal}{Journal of Physics A: Mathematical
  and Theoretical} \textbf{\bibinfo{volume}{42}}(\bibinfo{number}{50}),
  \bibinfo{pages}{504003} (\bibinfo{year}{2009}).

\bibitem{herviou}
\bibinfo{author}{\bibfnamefont{L.}~\bibnamefont{Herviou}},
  \bibinfo{author}{\bibfnamefont{N.}~\bibnamefont{Regnault}}, \bibnamefont{and}
  \bibinfo{author}{\bibfnamefont{J.~H.} \bibnamefont{Bardarson}},
  \emph{\bibinfo{title}{{Entanglement spectrum and symmetries in non-Hermitian
  fermionic non-interacting models}}}, \bibinfo{journal}{SciPost Phys.}
  \textbf{\bibinfo{volume}{7}}, \bibinfo{pages}{69} (\bibinfo{year}{2019}).

\bibitem{chen20}
\bibinfo{author}{\bibfnamefont{L.-M.} \bibnamefont{Chen}},
  \bibinfo{author}{\bibfnamefont{S.~A.} \bibnamefont{Chen}}, \bibnamefont{and}
  \bibinfo{author}{\bibfnamefont{P.}~\bibnamefont{Ye}},
  \emph{\bibinfo{title}{Entanglement, non-hermiticity and duality}}
  \bibinfo{note}{{a}rXiv:2009.00546}.

\bibitem{chang20}
\bibinfo{author}{\bibfnamefont{P.-Y.} \bibnamefont{Chang}},
  \bibinfo{author}{\bibfnamefont{J.-S.} \bibnamefont{You}},
  \bibinfo{author}{\bibfnamefont{X.}~\bibnamefont{Wen}}, \bibnamefont{and}
  \bibinfo{author}{\bibfnamefont{S.}~\bibnamefont{Ryu}},
  \emph{\bibinfo{title}{Entanglement spectrum and entropy in topological
  non-hermitian systems and nonunitary conformal field theory}},
  \bibinfo{journal}{Phys. Rev. Research} \textbf{\bibinfo{volume}{2}},
  \bibinfo{pages}{033069} (\bibinfo{year}{2020}).

\bibitem{maity}
\bibinfo{author}{\bibfnamefont{S.}~\bibnamefont{Maity}},
  \bibinfo{author}{\bibfnamefont{S.}~\bibnamefont{Bandyopadhyay}},
  \bibinfo{author}{\bibfnamefont{S.}~\bibnamefont{Bhattacharjee}},
  \bibnamefont{and} \bibinfo{author}{\bibfnamefont{A.}~\bibnamefont{Dutta}},
  \emph{\bibinfo{title}{Growth of mutual information in a quenched
  one-dimensional open quantum many-body system}}, \bibinfo{journal}{Phys. Rev.
  B} \textbf{\bibinfo{volume}{101}}, \bibinfo{pages}{180301}
  (\bibinfo{year}{2020}).

\bibitem{graefe2008}
\bibinfo{author}{\bibfnamefont{E.~M.} \bibnamefont{Graefe}},
  \bibinfo{author}{\bibfnamefont{H.~J.} \bibnamefont{Korsch}},
  \bibnamefont{and} \bibinfo{author}{\bibfnamefont{A.~E.}
  \bibnamefont{Niederle}}, \emph{\bibinfo{title}{Mean-field dynamics of a
  non-hermitian bose-hubbard dimer}}, \bibinfo{journal}{Phys. Rev. Lett.}
  \textbf{\bibinfo{volume}{101}}, \bibinfo{pages}{150408}
  (\bibinfo{year}{2008}).

\bibitem{giamarchi}
\bibinfo{author}{\bibfnamefont{T.}~\bibnamefont{Giamarchi}},
  \emph{\bibinfo{title}{Quantum Physics in One Dimension}}
  (\bibinfo{publisher}{Oxford University Press}, \bibinfo{address}{Oxford},
  \bibinfo{year}{2004}).

\bibitem{groha}
\bibinfo{author}{\bibfnamefont{S.}~\bibnamefont{Groha}},
  \bibinfo{author}{\bibfnamefont{F.~H.~L.} \bibnamefont{Essler}},
  \bibnamefont{and}
  \bibinfo{author}{\bibfnamefont{P.}~\bibnamefont{Calabrese}},
  \emph{\bibinfo{title}{{Full counting statistics in the transverse field Ising
  chain}}}, \bibinfo{journal}{SciPost Phys.} \textbf{\bibinfo{volume}{4}},
  \bibinfo{pages}{43} (\bibinfo{year}{2018}).

\bibitem{fagotti}
\bibinfo{author}{\bibfnamefont{M.}~\bibnamefont{Fagotti}} \bibnamefont{and}
  \bibinfo{author}{\bibfnamefont{P.}~\bibnamefont{Calabrese}},
  \emph{\bibinfo{title}{Evolution of entanglement entropy following a quantum
  quench: Analytic results for the $xy$ chain in a transverse magnetic field}},
  \bibinfo{journal}{Phys. Rev. A} \textbf{\bibinfo{volume}{78}},
  \bibinfo{pages}{010306} (\bibinfo{year}{2008}).

\bibitem{Alba2017}
\bibinfo{author}{\bibfnamefont{V.}~\bibnamefont{Alba}} \bibnamefont{and}
  \bibinfo{author}{\bibfnamefont{P.}~\bibnamefont{Calabrese}},
  \emph{\bibinfo{title}{Entanglement and thermodynamics after a quantum quench
  in integrable systems}}, \bibinfo{journal}{Proceedings of the National
  Academy of Sciences} \textbf{\bibinfo{volume}{114}}(\bibinfo{number}{30}),
  \bibinfo{pages}{7947} (\bibinfo{year}{2017}).

\bibitem{iuccisinegordon}
\bibinfo{author}{\bibfnamefont{A.}~\bibnamefont{Iucci}} \bibnamefont{and}
  \bibinfo{author}{\bibfnamefont{M.~A.} \bibnamefont{Cazalilla}},
  \emph{\bibinfo{title}{Quantum quench dynamics of the sine-gordon model in
  some solvable limits}}, \bibinfo{journal}{New J. Phys.}
  \textbf{\bibinfo{volume}{12}}, \bibinfo{pages}{055019}
  (\bibinfo{year}{2010}).

\end{thebibliography}

\appendix

\section{Derivation of the Green's function in one wavenumber channel\label{app1}}

The Green's function of the SSH model is defined by Eq. (6) of the main text. In the equation, the Green's matrix corresponding to a single wavenumber channel $k$ is given by
\begin{gather}
\Gv_k(t)=\left[\begin{array}{cc}\langle \Psi(t) | c_{Ak}c_{Ak}^{+}|\Psi(t)\rangle & \langle \Psi(t) | c_{Ak}c_{Bk}^{+}|\Psi(t)\rangle \\
\langle \Psi(t) | c_{Bk}c_{Ak}^{+}|\Psi(t)\rangle & \langle \Psi(t) | c_{Bk}c_{Bk}^{+}|\Psi(t)\rangle \end{array}\right].
\end{gather}
In this appendix, we consider one wavenumber channel only and the index $k$ will be dropped henceforth. The wavefunction 
$|\Psi(t)\rangle$ is also regarded as the wavefunction of the single mode $k$. Using the whole wavefunction, the Green's 
matrix would be proportional to the product of norm squares in all the other modes which are not necessarily unity due to 
the non-hermitian dynamics. However, in a latter phase of the calculation, the Green's matrix would be normalized by this overall factor.

Using the eigenbasis of the initial Hamiltonian, the Green's matrix is rewritten as
\begin{gather}
\Gv(t)= \Uv \gv(t) \Uv^{+},
\label{eq:rotation}
\end{gather}
where
\begin{gather}
\mathbf{U}=\frac{1}{\sqrt{2}}\left[\begin{array}{cc} 1 & e^{\frac{ika}{2}} \\ -e^{-\frac{ika}{2}} & 1\end{array} \right]
\end{gather}
is the unitary transformation between the sublattice basis and the initial Hamiltonian eigenbasis and
\begin{gather}
\gv(t)=\left[\begin{array}{cc}\langle \Psi(t) | d_{+}d_{+}^{+}|\Psi(t)\rangle & \langle \Psi(t) | d_{+}d_{-}^{+}|\Psi(t)\rangle \\
\langle \Psi(t) | d_{-}d_{+}^{+}|\Psi(t)\rangle & \langle \Psi(t) | d_{-}d_{-}^{+}|\Psi(t)\rangle \end{array}\right]
\end{gather}
with $d_\pm$ being the annihilation operators in the upper and lower band. The initial Hamiltonian is diagonalized by $d_\pm$ as given in Eq. (2) of the main text.
After the quantum quench, the Hamiltonian  becomes
\begin{gather}
H=\varepsilon(d_{+}^{+}d_{+}-d_{-}^{+}d_{-})  + 
h \left(\alpha_+ d_{+}^{+}d_{-}  +  \alpha_- d_{-}^{+}d_{+}\right),
\end{gather}
where $\varepsilon = 2J\cos \left(\frac{ka}{2}\right)$ and $\alpha_\pm = i e^{\pm i \frac{ka}{2} } \left(1 \pm \sin\left(\frac{ka}{2}\right)\right)$.
We evaluate the time dependence of the matrix elements $g_{ss'}(t)=\langle\Psi(t)|d_sd_{s'}^{+}|\Psi(t)\rangle$ with $s,s' = \pm$ using the non-Hermitian Schr\"odinger equation leading to
\begin{gather}
\partial_t g_{ss'}(t)=i\langle\Psi(t)| H^{+}d_sd_{s'}^{+}-d_sd_{s'}^{+}H|\Psi(t)\rangle
\label{eom1}
\end{gather}
(all operators acting to the right). Since $H\neq H^+$, both commutators and anticommutators arise in Eq. \eqref{eom1}. 
The "commutators" emerging in the differential equations are obtained as
\begin{subequations}
\begin{gather}
H^{+}d_+d_+^{+} - d_+d_+^{+} H = h \alpha_- d_+d_-^+ - h \alpha_-^* d_-d_+^+   \\
H^{+}d_-d_-^{+} - d_-d_-^{+} H = h \alpha_+ d_-d_+^+ - h \alpha_+^* d_+d_-^+ \\
H^{+}d_+d_-^{+} - d_+d_-^{+} H = -\left(H^{+}d_-d_+^{+} - d_-d_+^{+} H\right)^{+} = \nonumber \\
=h\left(\alpha_+ d_+d_+^+ - \alpha_-^* d_-d_-^+  +(\alpha_-^* -\alpha_+)d_+d_+^+d_-d_-^+   \right)  -\nonumber\\
-2\varepsilon d_+d_-^+ .
\end{gather}
\end{subequations}
The equations show that the first derivatives are in linear relation with the coefficients $g_{ss'}(t)$ themselves. In addition, due to the anticommutator term in Eq. \eqref{eom1},
they are coupled to the function
$D(t)=\langle\Psi(t)| d_+d_+^+d_-d_-^+|\Psi(t)\rangle$
which, however, is constant zero, since $H^{+}d_+d_+^+d_-d_-^+ - d_+d_+^+d_-d_-^+ H = 0$
and $D(0)=0$ in the initial state.

To summarize, the dynamics of the elements of $\gv(t)$ are given by
\begin{subequations}
\begin{gather}
\frac{1}{h}\partial_t g_{++}(t) = i\alpha_- g_{+-}(t) - i\alpha_-^* g_{-+}(t), \\
\frac{1}{h}\partial_t g_{--}(t) = i\alpha_+ g_{-+}(t) - i\alpha_+^* g_{+-}(t), \\
\frac{1}{h}\partial_t g_{+-}(t) = -i\frac{2\varepsilon}{h} g_{+-}(t) + i\alpha_+ g_{++}(t) - i\alpha_{-}^* g_{--}(t), \\
\frac{1}{h}\partial_t g_{-+}(t) = i\frac{2\varepsilon}{h} g_{-+}(t) - i\alpha_+^* g_{++}(t) + i\alpha_{-} g_{--}(t) 
\end{gather}
\end{subequations}
with the initial conditions of $g_{++}(0)=1$ and $g_{--}(0)=g_{+-}(0)=g_{-+}(0)=0$.
The differential equations are solved by
\begin{subequations}
\begin{gather}
g_{++}(t)= 1 + A(t),\hspace*{5mm}g_{--}(t)= C\,A(t), \\
g_{+-}(t)=g_{-+}(t)^* = -e^{i\frac{ka}{2}}\sqrt{C} B(t) + 
ie^{i\frac{ka}{2}}\sqrt{C}\frac{2J}{h} A(t),
\end{gather}
\end{subequations}
where we defined 
\begin{subequations}
\begin{gather}
A(t)=\frac{h^2}{2E_0^2}\left(1-\cos(2Et)\right), \\
B(t) = \frac{h}{2E_0}\sin(2Et), \\
C = \frac{1+\sin(ka/2)}{1 - \sin(ka/2)}
\end{gather}
\label{eq:ABC}
\end{subequations}
with $E=E_0\cos(ka/2)$ being the spectrum of the non-Hermitian Hamiltonian and $E_0 = \sqrt{4J^2-h^2}$.

Rotating back to the sublattice basis based on \eqref{eq:rotation}, we obtain
\begin{widetext}
\begin{gather}
\Gv(t)= \frac{1}{2} \left[\begin{array}{cc} 1 + (1+C) A(t) - 2\sqrt{C}B(t) & e^{i\frac{ka}{2}} \left(-1-A(t)+CA(t) + 2i\sqrt{C}\frac{2J}{h}A(t)\right) \\
e^{-i\frac{ka}{2}} \left(-1-A(t)+CA(t) - 2i\sqrt{C}\frac{2J}{h}A(t)\right) & 1 + (1+C) A(t) + 2\sqrt{C}B(t) \end{array}\right].
\end{gather}
\end{widetext}
The norm of the wavefunction changes with time due to the non-hermitian dynamics\cite{daley,ashida18}. 
The norm square $\mathcal{N}(t)=\langle \Psi(t)|\Psi(t) \rangle$ obeys the differential equation 
$\partial_t \mathcal{N}(t) = i\langle \Psi(t)|H^{+}-H|\Psi(t)\rangle$
and $\mathcal{N}(0)=1$ which is solved by
\begin{gather}
\mathcal{N}(t)=1 + (1+C)A(t),
\end{gather}
where $A(t)$ and $C$ are defined in Eq. \eqref{eq:ABC}.
The Green's function is normalized with the norm, yielding Eq. (7) in the main text.

We note that this is the normalization step where the norm square of the other wavenumber channels would also cancel out if they had been considered in the Green's function by its definition.

\section{Lindblad time evolution of the open SSH model}
In the main text, we consider the time evolution of the SSH model governed by a non-Hermitian Hamiltonian as given 
in Eq. (3) of the main text. The non-Hermitian Hamiltonian may be derived from a Lindblad equation 
with omitting the recycling term\cite{daley,carmichael} and quantum jump processes. The corresponding Lindblad equation is determined by the jump operators
\begin{gather}
L_{Aj}=c_{Aj},\qquad\qquad L_{Bj}=c_{Bj}^+
\end{gather}
and the Lindblad equation reads
\begin{gather}
\partial_t \rho = -i\left[H_r,\rho\right] + \mathcal{D}[\rho],
\label{eq:lindblad}
\end{gather}
where $\rho$ is the density matrix and $H_r$ is the Hamiltonian Eq. (3) of the main text without the imaginary site energy terms. The dissipation term is given as
\begin{gather}
\mathcal{D}[\rho] = h\sum_{j,X=A,B} \left(2L_{Xj}\rho L_{Xj}^+ - L_{Xj}^+L_{Xj}\rho - \rho L_{Xj}^+L_{Xj} \right).
\end{gather}
The Lindblad equation is completely decoupled with respect to the wavenumbers $k$.
For a given wavenumber, the Fock space is 4-dimensional spanned by
$|0,0\rangle$, $|0,1\rangle$, $|1,0\rangle$ and $|1,1\rangle$
where the first (second) number is the occupation number of the A (B) sublattice.
In this basis, the creation and annihilation operators are represented by
\begin{gather}
\uu{c}_A =\left[\begin{array}{cccc} 0 & 0 & 1 & 0 \\ 0 & 0 & 0 & 1 \\ 0 & 0 & 0 & 0 \\ 0 & 0 & 0 & 0 \end{array}\right] ,
\qquad \uu{c}_B =\left[\begin{array}{cccc} 0 & 1 & 0 & 0 \\ 0 & 0 & 0 & 0 \\ 0 & 0 & 0 & -1 \\ 0 & 0 & 0 & 0 \end{array}\right] 
\end{gather}
and the initial density matrix is represented by
\begin{gather}
\uu{\rho}_0=\frac{1}{2}\left[\begin{array}{cccc} 0 & 0 & 0 & 0 \\ 0 & 1 & e^{-i\frac{ka}{2}} & 0 \\ 0 & e^{i\frac{ka}{2}} & 1 & 0 \\ 0 & 0 & 0 & 0 \end{array}\right],
\end{gather}
which describes the same half-filled, initial state as in the case of the non-Hermitian dynamics.

In general, the $4\times 4$ density matrix $\uu{\rho}(t)$ has 4 real-valued and 6 complex-valued independent entries. However, the specific dynamics determined by the Lindblad 
equation \eqref{eq:lindblad} and the initial state ensures that the density matrix has the simpler form of
\begin{gather}
\uu{\rho}(t)= \alpha(t)\uu{m}_2 + \beta(t) \uu{m}_{12} + \gamma(t)\uu{m}_3+ \uu{q}(z(t)),
\end{gather}
where
\begin{gather}
\uu{m}_2 = \left[\begin{array}{cccc} 0 & 0 & 0 & 0 \\ 0 & 1 & 0 & 0 \\ 0 & 0 & 0 & 0 \\ 0 & 0 & 0 & 0 \end{array}\right], \qquad
\uu{m}_{12} = \left[\begin{array}{cccc} \frac{1}{2} & 0 & 0 & 0 \\ 0 & -1 & 0 & 0 \\ 0 & 0 & 0 & 0 \\ 0 & 0 & 0 & \frac{1}{2} \end{array}\right], \nonumber \\
\uu{m}_3 = \left[\begin{array}{cccc} -1 & 0 & 0 & 0 \\ 0 & 1 & 0 & 0 \\ 0 & 0 & 1 & 0 \\ 0 & 0 & 0 & -1 \end{array}\right], \qquad
\uu{q}(z)= \left[\begin{array}{cccc} 0 & 0 & 0 & 0 \\ 0 & 0 & z^* & 0 \\ 0 & z & 0 & 0 \\ 0 & 0 & 0 & 0 \end{array}\right]
\end{gather}
are eigenmatrices of the dissipation term of the Lindblad equation. In the formula, $\alpha(t)$, $\beta(t)$ and $\gamma(t)$ 
are real-valued functions while $z(t)$ is complex-valued. Since the trace of the density matrix is unity, we obtain $\alpha(t)=1$. Based on the Lindblad equation, the following differential equations are derived.

\begin{subequations}
\begin{gather}
\dot{\beta}=2\left(if^*z(t)-ifz^*(t)\right) - 2h \beta(t), \\
\dot{\gamma}=\left(if^*z(t)-ifz^*(t)\right) - 4h \gamma(t), \\
\dot{z}=-if+if\beta(t) - 2hz(t)
\end{gather}
\end{subequations}
with the initial conditions of $\beta(t=0) = 1$, $\gamma(t=0) = \frac{1}{2}$ and $z(t=0) = \frac{1}{2}e^{i\frac{ka}{2}}$. In the formulas, $f = \left(-J-\frac{h}{2}\right) + \left(-J+\frac{h}{2}\right)e^{ika}$. The differential equations are solved by
\begin{widetext}
\begin{subequations}
\begin{gather}
\beta(t)=\frac{g^2}{1+g^2} + e^{-2ht}\left[\frac{\cos(2|f|t)}{1+g^2} + \left(\sin\left(\frac{ka}{2}\right) - \frac{g^2}{g^2+1}\right)\frac{\sin(2|f|t)}{g}\right], \\
\gamma(t) = \frac{1}{4}\frac{g^2}{1+g^2}+e^{-4ht}\left[\frac{1}{2}-\frac{1}{4}\frac{g^2}{1+g^2}-\frac{1}{2}\frac{\sin\left(\frac{ka}{2}\right)}{1+g^2}\right]
+ e^{-2ht}\left[\cos(2|f|t)\frac{1}{2}\frac{\sin\left(\frac{ka}{2}\right)}{1+g^2}-\sin(2|f|t)\frac{g}{2}\frac{1-\sin\left(\frac{ka}{2}\right)}{1+g^2}\right], \\
z(t)=h\frac{V(t)+iW(t)}{2if^*},
\end{gather}
\label{eq:sollindblad}
\end{subequations}
where $g=|f|/h=\sqrt{h^2+E_0^2\cos^2(ka/2)}/h$ and
\begin{subequations}
\begin{gather}
V(t)=\frac{g^2}{1+g^2} + g e^{-2ht}\left[-\frac{\sin(2|f|t)}{1+g^2} + \left(\sin\left(\frac{ka}{2}\right) - \frac{g^2}{g^2+1}\right)\frac{\cos(2|f|t)}{g}\right] \\
W(t)=-\frac{2J}{h}e^{-2ht}\cos\left(\frac{ka}{2}\right).
\end{gather}
\end{subequations}
\end{widetext}
The functions in Eqs. \eqref{eq:sollindblad} determine the time evolution of the density matrix in the wavenumber channel $k$. 

\begin{figure}[h!]
\centering
a)\includegraphics[width=7cm]{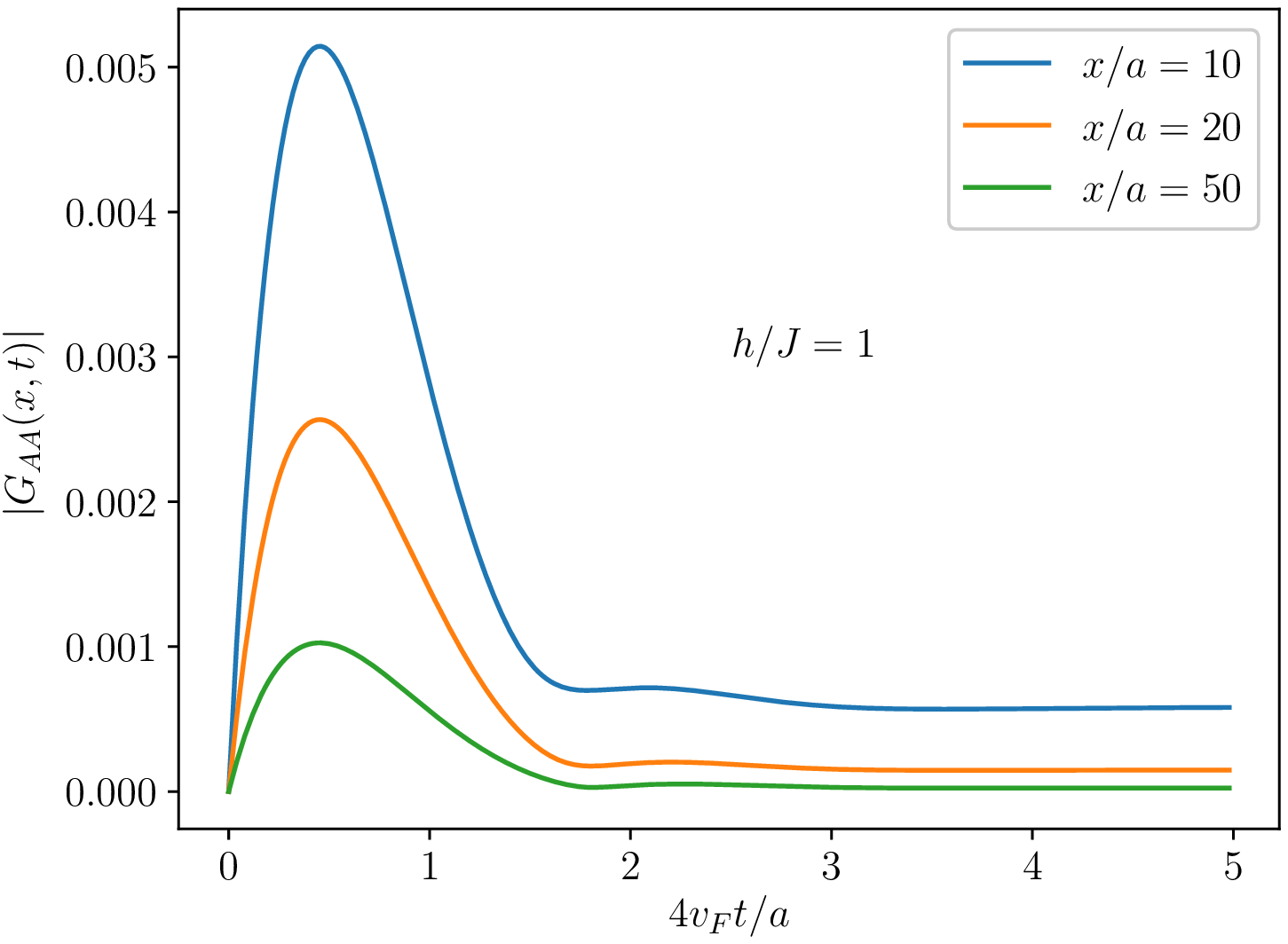}
b)\includegraphics[width=7cm]{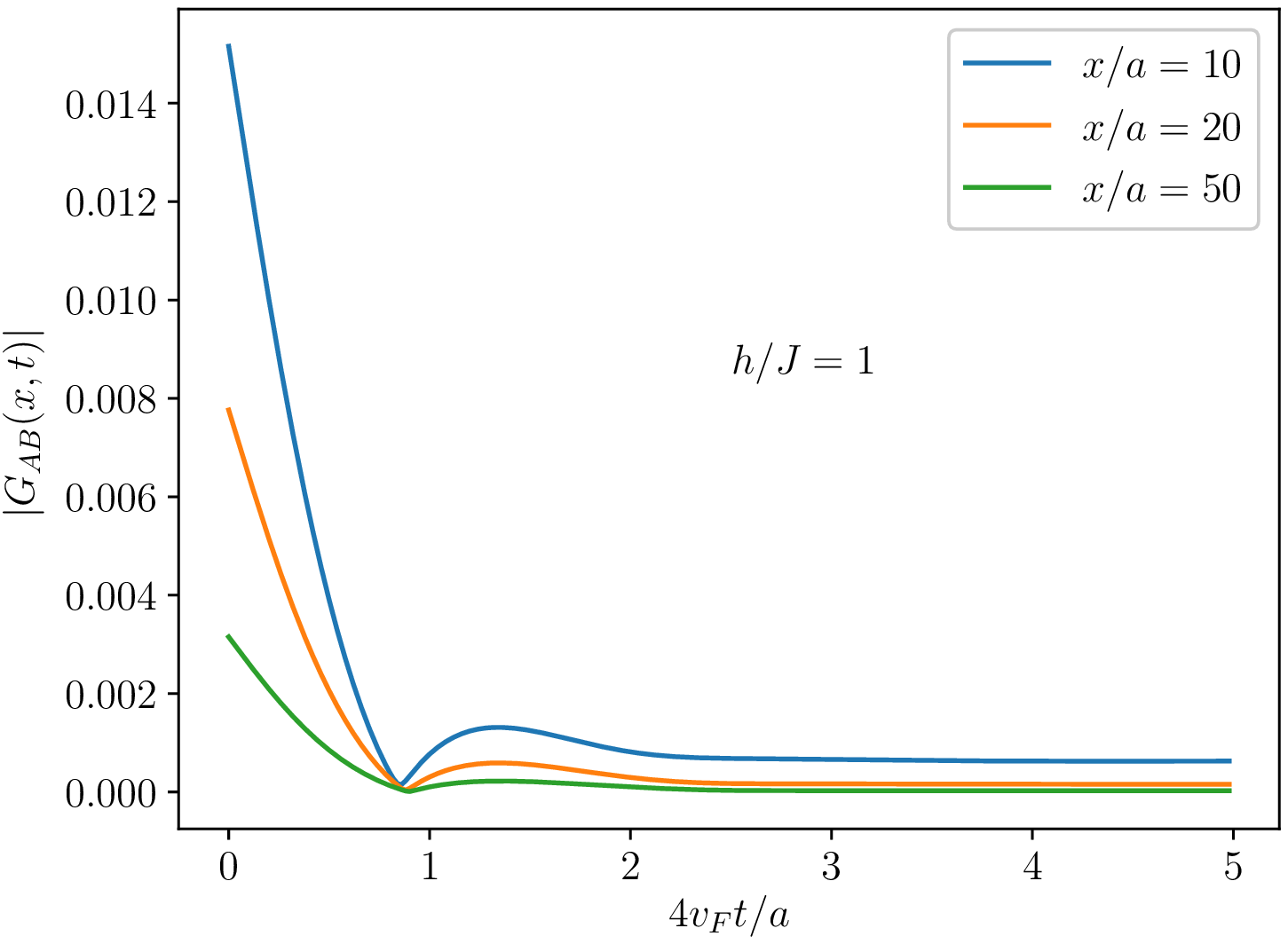}
\caption{The post-quench time-dependence of diagonal (a)
and off-diagonal (b) Green's matrix elements governed by the Lindblad equation. Both matrix elements exhibit exponential decay after transients. No signal of supersonic modes can be observed.}
\label{fig:greenlindblad}
\end{figure}

Based on the density matrix, the Green's function is calculated as
\begin{gather}
G_{XX'}(x_{j'}-x_j,t)=\mathrm{Tr}\left[\rho(t)c_{Xj}c_{X'j'}^{+}\right]=\nonumber \\
=\frac{1}{N}\sum_k e^{ik(x_{j'}-x_{j})}G_{XX',k}(t),
\end{gather}
where $X,X'=A,B$ and $N$ is the number of unit cells. The matrix $\Gv_k(t)$ corresponding to one wavenumber channel is obtained as
\begin{gather}
\Gv_k(t) = \left[\begin{array}{cc} 1-\frac{\beta(t)}{2} & -z(t) \\ -z^*(t) & \frac{\beta(t)}{2}\end{array}\right],
\end{gather}
which is the Lindbladian counterpart of Eq. (7) in the main text.

The Fourier transformation is evaluated numerically and is plotted in Fig. \ref{fig:greenlindblad}. The main observation is that there are no supersonic modes in the time evolution. Initial correlations are suppressed with the function $e^{-2ht}$. For comparison, see Fig. 2 in the main text.

In the steady state, we find
\begin{gather}
\Gv_k^\infty = \frac{1}{2}\left[\begin{array}{cc} 1 + \frac{h^2 }{h^2+|f|^2} & \frac{ihf}{h^2+|f|^2} \\ \frac{-ihf^*}{h^2+|f|^2} & 1 - \frac{h^2}{h^2+|f|^2}\end{array}\right],
\end{gather}
which can be Fourier transformed numerically only. Numerical results indicate that the diagonal elements show a spatial decay as $x^{-2}$ while off-diagonal elements decrease as $x^{-1}$.



\section{Quench from hermitian gapless to non-hermitian gapped SSH model}

\jav{In this section, we present the Green's function in a SSH model which is a variant of the one presented in the main text. 
In this model, the initial state and the Hamiltonian are considered to be the same as in the main text. At $t=0$, we suddenly switch on alternating hopping and 
imaginary, alternating site energy as
\begin{gather}
H=H_0 + \sum_{j}\left[ih \left(c_{Aj}^+ c_{Aj} - c_{Bj}^+c_{Bj}\right) + \right. \nonumber \\
\left. + \delta\left(c_{Aj}^+c_{Bj-1} + c_{Bj-1}^+ c_{Aj}\right) - \delta\left(c_{Bj}^+c_{Aj} + c_{Bj}^+c_{Aj}\right) \right]
\label{eq:fullham2}
\end{gather}
which is the generalization of Eq. \eqref{eq:fullham} where we had $\delta=h/2$. If $\delta$ differs from $h/2$, the two energy 
bands are still real-valued but separated by a gap. The normalized Green's function of this model is obtained as
\begin{widetext}
\begin{subequations}
\begin{gather}
\tilde{\Gv}_k(t) = \Gv_k(0) + \frac{1}{\mathcal{N}_k(t)}\left[ \begin{array}{cc} -\beta_k B_k(t) & e^{i\frac{ka}{2}} \frac{A_k(t)}{\beta_{-k}}\left(\beta_k + i\frac{\varepsilon(k)}{h}\right) \\
e^{-i\frac{ka}{2}} \frac{A_k(t)}{\beta_{-k}}\left(\beta_k - i\frac{\varepsilon(k)}{h}\right) & \beta_k B_k(t) \end{array}\right] \\
\beta_k = 1 + \frac{2\delta}{h}\sin\left(\frac{ka}{2}\right) \qquad\qquad \varepsilon(k) = 2J\cos\left(\frac{ka}{2}\right)\\
A_k(t) = \frac{h^2}{2E(k)^2}\left(1-\frac{4\delta^2}{h^2}\sin^2\left(\frac{ka}{2}\right)\right) \left(1 - \cos(2E(k)t)\right) \qquad\qquad
B_k(t) = \frac{h}{2E(k)}\sin(2E(k)t) \\
E(k) = \sqrt{(4J^2-4\delta^2)\cos^2\left(\frac{ka}{2}\right) - h^2 + 4\delta^2} \\
\mathcal{N}_k(t) = 1 + \left(1 + \frac{\beta_k}{\beta_{-k}}\right) A_k(t)
\end{gather}
\end{subequations}
\end{widetext}
where $\Gv_k(0)$ is the initial Green's matrix as given in Eq. \eqref{eq:g0}.
By numerically evaluating the Fourier transform, we find that the real space Green's function  decays as $1/x^2$ in the long time limit.}

\section{Quench from gapped hermitian SSH model to EP}

\jav{In this section, we focus on the Green's function after a quench from a gapped SSH model to an EP.
The alternating hopping is already present in the initial state, which represents the fully gapped SSH model. Then, we quench
the alternating imaginary site energies on top of the SSH Hamiltonian, such, that the non-hermitian Hamiltonian contains an EP. The initial Hamiltonian reads
\begin{gather}
H_0 = \sum_{j}\left[\left(-J+\frac{h}{2} \right) \left(c_{Aj}^{+}c_{Bj-1}+c_{Bj-1}^+c_{Aj}\right) + \right. \nonumber \\
\left. + \left(-J-\frac{h}{2}\right)\left(c_{Aj}^{+}c_{Bj} + c_{Bj}^{+}c_{Aj}\right)\right]
\end{gather}
whose energy bands,
\begin{gather}
\varepsilon(k) = \sqrt{2J^2 +\frac{h^2}{2} + \left(2J^2 - \frac{h^2}{2}\right) \cos\left(ka\right)}
\end{gather}
and $-\varepsilon(k)$, are separated by a gap of size $|h|$ for $2|J|>|h|$. The initial state is considered to be the ground state of the half-filled system. At $t=0$, the imaginary, 
alternating site energy is switched on and the time-evolution is driven by the Hamiltonian
\begin{gather}
H=H_0 + ih\sum_{j}\left(c_{Aj}^+c_{Aj} - c_{Bj}^+c_{Bj}\right)
\end{gather}
which is the same as in Eq. (3) of the main text. Following the same method as in Appendix \ref{app1}, the normalized Green's function is obtained as
\begin{gather}
\tilde{\Gv}_k(t) = \Gv_k(0) + \frac{1}{\mathcal{N}_k(t)}\left[\begin{array}{cc}
 - B_k(t) & C_k(t) \\
C^{*}_k(t) & B_k(t) \end{array}\right]
\end{gather}
where
\begin{gather}
B_k(t)=\frac{h}{2E(k)}\sin(2E(k)t)\\
C_k(t)=- \frac{hf(k)}{2\varepsilon(k)E(k)^2}\left(h + i\varepsilon(k)\right)\left(1-\cos(2E(k)t)\right) \\
\mathcal{N}_k(t) = 1+\frac{h^2}{E(k)^{2}}\left(1-\cos(2E(k)t)\right) \\
f(k) = \left(-J+\frac{h}{2}\right)e^{ika} -J -\frac{h}{2},
\end{gather}
$E(k) =  \sqrt{4J^2-h^2}\cos({ka}/{2})$ and
\begin{gather}
\Gv_{k}(0)=\frac{1}{2}\left[\begin{array}{cc}
1 & \frac{f(k)}{\varepsilon(k)} \\
\frac{f^*(k)}{\varepsilon(k)} & 1 \end{array}\right]
\end{gather}
is the initial Green's matrix.}

\jav{The Fourier transform of the initial off-diagonal Green's function decays exponentially, as expected from the finite gap in the spectrum, while the diagonal components 
is local as $\delta_{x,0}$.
In the infinite time limit, one can take the time average of the Green's function to obtain
\begin{gather}
\tilde{\Gv}_k^\infty = \Gv_k(0) + \left[\begin{array}{cc}
 0 & D_k \\
D^{*}_k & 0 \end{array}\right]
\end{gather}
where
\begin{gather}
D_k=- \frac{hf(k)}{2\varepsilon(k)}\frac{h + i\varepsilon(k)}{\sqrt{E^2(k)+2h^2}(E(k)+\sqrt{E^2(k)+2h^2)}}
\end{gather}
The Fourier transformation of the  off-diagonal component to real space was investigated numerically, and was found to decay as $1/x^2$. This is somewhat surprising,
because in the hermitian case, a gapped to gapless quench induces correlations decaying exponentially\cite{iuccisinegordon}. In the current case, this power law decay
can then be attributed to the non-hermitian nature of the quench, namely that the gapless point is an EP.}

\end{document}